\documentclass[twocolumn,aps,showpacs]{revtex4}
\usepackage{amsmath}
\usepackage{graphics,graphicx}
\usepackage{color}
\usepackage[normalem]{ulem}
\def\delete#1{}
\def\add#1{\textcolor[rgb]{0,0,0}{#1}}

\begin{document}
\title{Diffraction-free nonevanescent nano-beams using the Fresnel-waveguide concept}
\author{M.I.\ Mechler\textsuperscript{1}, Z.\ Tibai\textsuperscript{2}, S.V.\ Kukhlevsky\textsuperscript{2}}
\affiliation{\textsuperscript{1}High-Field Terahertz Research Group, MTA-PTE, Ifj\'us\'ag \'u.\ 6, 7624 P\'ecs, Hungary}
\affiliation{\textsuperscript{2}Institute of Physics, University of P\'ecs, Ifj\'us\'ag \'u.\ 6,  7624 P\'ecs, Hungary}
\begin{abstract}
Strong intensity attenuation limits the use of conventional diffraction-free optical elements. We show a possible solution to the exponential intensity attenuation limiting the use of Fresnel-type diffraction-free nanometer-scale optics by using materials with appropriately chosen refractive index. Such beams may be applied for technical and physical problems.
\end{abstract}
\maketitle
\section{Introduction}
Resolving micro- and nanoscopic objects is a challenging problem and it has been an important field of research in the last decades. Although a simple microscope is sufficient to magnify microscopic images, its resolution capability is confined by the Abbe diffraction limit:
\begin{equation}
d=\frac{\lambda}{2n\sin(\theta)},\label{eq:Abbe}
\end{equation}
where $\lambda$ is the wavelength, $n$ is the refractive index while the angle of incidence is denoted by $\theta$. Therefore diffraction is an ultimate barrier in traditional microscopy.

Many solutions have been proposed to overcome this limit, that is, to eliminate its effect, such as photoactivatable localization microscopy\cite{Betz2006} (PALM), stimulated emission depletion\cite{Hell1994} (STED), surface plasmon polaritons\cite{SPPhivs} (SPPs), various nanostructures such as the recently discovered microspheres\cite{microspheres}, or the diffraction-free wave solutions \cite{DFB,Gori,Her,Haf,Ru,Cha1,Saa,Bou,Por1,Baj}. \add{Ultrashort diffraction-free pulses can also be produced \cite{kuknyit,kukepl,KukOComm}.} Diffraction-free light beams can be defined as beams with no diffractive broadening throughout the free-space propagation. This can be expressed as follows: for all $x$, $y$ and $z$ coordinates the average transverse components $\mathbf{I}_T=\overline{\mathbf{S}}_T$ of the Poynting vector $\mathbf{S}_T$ of the beam must fulfill the following condition \cite{dlessdef}:
\begin{equation}
\nabla\cdot\mathbf{I}_T=0
\end{equation}
In \cite{KukOComm} we described a diffraction-free arrangement that consists of an array of slits illuminated by TM-waves with periodically altering phase; however, it was also pointed out that this diffraction-free feature implies an exponential decrease in the intensity, i.e.\ behind the slit array evanescent beams can be observed.

Evanescent beams are described through an exponential term:
\begin{equation}
U(x,y,z)=F(x,y,z)\cdot\exp(i\beta z),
\end{equation}
where $z$ is the coordinate along the propagation direction, and $\beta$ is the wavevector component in the direction of propagation. If $\beta$ is purely imaginary, the beam becomes evanescent; however, if this wavevector component remains real, propagating waves are generated.

This was the basic idea of Ruschin and Leizer \cite{Ruschin}. In their paper they concentrated on Bessel-type beams which are generally treated as propagating waves. They focused on the problem of generating evanescent Bessel-beams, therefore in their work they used electromagnetic solutions of the form
\begin{equation}
U_m(r,\phi,z)=\exp(i\phi m)J_m(\alpha r)\exp(-\beta'z),
\end{equation}
where $\beta'=\sqrt{\alpha^2-n^2k_0^2}$ is the propagation constant, $(r,\phi,z)$ are cylindrical coordinates, $J_m$ is the Bessel function. Whenever $\beta'$ is real, i.e.\ $\alpha^2>n^2k_0^2$, this function decreases exponentially; however, if $\beta'$ is imaginary, the exponential part describes a propagating component. This can be achieved for example by changing the refractive index of the materials.

Another important result in this field was published by Yacob Ben-Aryeh\cite{BenAryeh}. This paper analyzes theoretically the microsphere-based field conversion from evanescent to propagating waves \cite{microspheres}. Information on the nanostructure of a nano-corrugated metallic thin film is included in the evanescent waves produced by the plane EM waves transmitted through the structure. This information can be extracted from the evanescent wave by microspheres located above the surface that collect and convert them into propagating waves. The conversion effect is also the consequence of the refractive index difference of the materials.

Nondiffracting beams have some special features which may be used in technical as well as physical applications such as imaging, particle acceleration or light motors\cite{Bouchal44, Bouchal66, Bouchal67, Bouchal72, Bouchal73, Bouchal74}. In these applications, however, the use of nonevanescent beams would be preferred.

In our paper we show that the production of propagating (nonevanescent) diffraction-free nano-beams using the Fresnel-waveguide concept is feasible. Our geometry is similar to a simple grating; however, the phase difference between the EM fields falling to the adjacent subwavelength slits and the near-field treatment of the problem distinguish our case from that of a simple grating. The paper is arranged as follows. In Section~\ref{sec:theory} we present the basic idea and the results of our theoretical analysis. Section~\ref{sec:numer} contains the results of our numerical simulations using a finite element analysis code. We present different geometries and conclude that critical refractive indices separating the evanescent and propagating regions depend on the transverse characteristic sizes of the geometries. Finally, in Section~\ref{sec:conclusion} conclusions are drawn and some final remarks are made.

\section{Theory}\label{sec:theory}
In \cite{KukOComm} we showed that it is possible to produce diffraction-free nano-beams in the subwavelength regime using the Fresnel-waveguide concept. Throughout our analysis we modeled an array of one-dimensional slits (Fig.~\ref{fig:origgeom}) of width $2a$ in a screen of thickness $b$ illuminated by TM-polarized plane waves of wavelength $\lambda$. The phase $\phi_m$ of the wave falling on the $m$-th slit of the array was adjusted as $\phi_m=m\cdot\pi$, $m=0,\pm1,\pm2\dots$ These phase jumps are very important as the phase difference of the fields distinguishes our problem from that of a grating. In our earlier paper we observed that the width of the central beam remains the same for great distances and the beam shape is sinusoid in the direction perpendicular to the propagation direction. However, we also presented that the intensity decreases exponentially. In this section we find a solution to this problem.

The beam leaving the slit array propagates in homogeneous material; in this case the propagation can be described by the Helmholtz equation and we may use the scalar field $U(\vec{r},t)$:
\begin{equation}
\left(\nabla^2+k^2\right)U(\vec{r},t)=0,\label{eq:Helmholtz}
\end{equation}
where $\nabla^2$ is the Laplacian and $k$ is the wavenumber. Owing to the geometry of the structure and the TM-polarization (z-invariance) of the incident light, in our case there is no $z$-dependency. We also disregard time dependency. Moreover, $U(x,y)$ can be split into two functions, $F(x)$ and $G(y)$, i.e.\ variables can be separated.

Now we assume an exponential form for the $y$-dependent part: $G(y)\sim\exp(ik_yy)$, where $k_y$ is the $y$-component of the wavevector (propagation constant). Substituting this into \eqref{eq:Helmholtz} we get:
\begin{equation}
\frac{\partial^2}{\partial x^2}F(x)=-(k^2-k_y^2)F(x).\label{eq:redHelm}
\end{equation}
Considering that a harmonic function is expected in the $x$ direction: $F(x)\sim\sin(\sqrt{k^2-k_y^2}x)$, by solving \eqref{eq:redHelm} for a beam with harmonic shape in the transverse direction and exponential function in the longitudinal direction, one may obtain the following function:
\begin{equation}
U(x,y)=F(x)\cdot G(y)\sim\sin\left(\frac{2\pi x}{4a}\right)\cdot\exp(ik_yy),\label{eq:ufg}
\end{equation}
where $k_y=\sqrt{\left(\frac{n2\pi}{\lambda_0}\right)^2-\left(\frac{2\pi}{4a}\right)^2}$, and $\lambda_0$ is the wavelength in vacuum. From this latter formula one may draw the conclusion that there are some situations (specific geometric dimensions and wavelengths) in which the otherwise evanescent waves become propagating, namely if $k_y^2<0$ the beam is evanescent while in the case of $k_y^2>0$ a propagating beam can be obtained. This leads to the following inequalities (note that all the quantities -- $n$, $a$, $\lambda_0$ -- are positive real values):
\begin{itemize}
\item evanescent beams are described by:
$$n<\frac{\lambda_0}{4a}$$
\item propagating beams are described by:
$$n>\frac{\lambda_0}{4a}$$
\end{itemize}
We also note that this formula was derived for the geometry in Figure~\ref{fig:origgeom} analyzed in our earlier paper \cite{KukOComm}. A more general formula can be written through the transverse characteristic dimension of the geometry, e.g.\ in our case $\Lambda=\left(\dfrac{2\pi}{4a}\right)^{-1}$, therefore
\begin{equation}
k_y=\sqrt{\left(\dfrac{n2\pi}{\lambda_0}\right)^2-\left(\dfrac{1}{\Lambda}\right)^2}.
\end{equation}

This deduction leads to the conclusion that in our geometry it is possible to achieve diffraction-free \emph{and} propagating waves by 1) increasing the refractive index 2) decreasing the wavelength or 3) increasing the transverse characteristic dimension of the geometry. In the following section we show some simulation results for various refractive indices and two different geometries (transverse characteritic sizes).
\section{Numerical simulations}\label{sec:numer}
In order to verify our result, we performed finite element analysis simulations for various geometries (transverse characteristic dimensions) and refractive indices. In the followings we present some results of our simulations.

\begin{figure}[tb]
\includegraphics[width=\columnwidth]{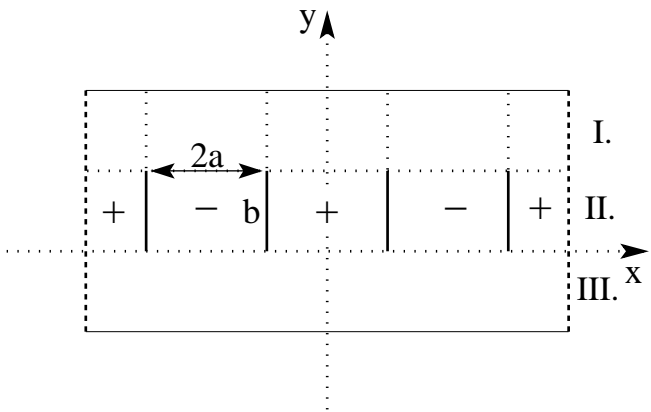}
\caption{\label{fig:origgeom}The original geometry used in \cite{KukOComm}. In the present article we used the following values: wavelength in vacuum $\lambda_0=800$~nm, slit width $2a=80$~nm, screen thickness $b=50$~nm. The "individual" slits are separated by perfectly conducting material. The phase of the EM field illuminating the slits changes periodically (in the figure it is indicated by the plus and minus characters)}
\end{figure}

The first simulated geometry was the same we analyzed in Section~\ref{sec:theory} (Fig.~\ref{fig:origgeom}).
In this case we used an array of slits of width $2a$ in a screen of thickness $b$. It is illuminated from region I
by TM polarized field that is perpendicular to the array of slits, and the phase of which was changed by $\pi$ in the adjacent slits, i.e.\ $\phi_m=m\cdot\pi$. In this case no distance was assumed between the individual slits, therefore this geometry is quite hypothetic. We will show a more realistic model hereinafter.

The aim of this paper was to find a solution to the exponential decrease of the wave leaving the slit (in region III). To prove that the solution proposed in Section~\ref{sec:theory} leads to nonevanescent (propagating) waves we plotted 
\begin{figure*}[tb]
\includegraphics[width=\textwidth]{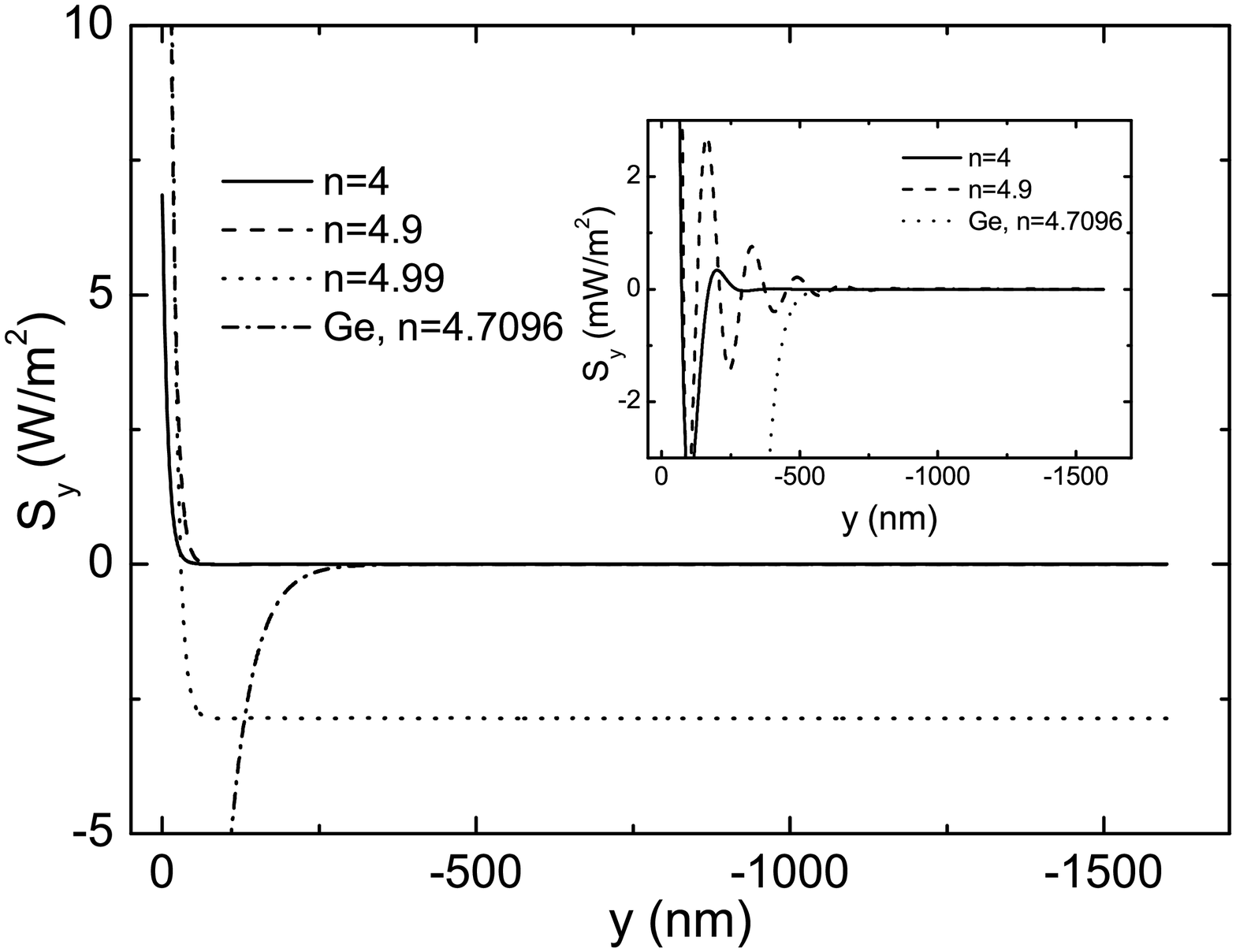}
\caption{\label{fig:Sy-origgeom}Energy flow on the propagation axis of the central slit calculated behind the slit array shown in Fig.~\ref{fig:origgeom} for various refractive indices. Slit width $2a=80$~nm, wavelength in vacuum $\lambda_0=800$~nm, screen thickness $b=50$~nm. For comparison we plotted the same quantity for a real material with high refractive index (germanium, $n\approx 4.7096$).
}
\end{figure*}
the energy flow $S_y$ along the $y$ axis behind the screen at $x=0$ for various refractive indices (Fig.~\ref{fig:Sy-origgeom}). For comparison the same quantity for a real material (germanium, $n=4.7096$ $@$ $\lambda=800$~nm \cite{Aspnes}) was also plotted. In this case the characteristic dimension of the geometry is $\Lambda=\left(\dfrac{4a}{2\pi}\right)$, therefore the critical refractive index of the geometry that separates the evanescent and propagating case can be formulated as $n=\frac{\lambda_0}{4a}$. Considering a wavelength of $\lambda_0=800$~nm and slit width $2a=80$~nm, the critical refractive index separating the evanescent and propagating cases is $n=5$. Indeed, Figure~\ref{fig:Sy-origgeom} confirms this value. Note that the transition from evanescent to propagating waves is accompanied by an oscillating effect, i.e.\ in the case of materials with refractive indices close to the critical value the $y$-component of the Poynting vector $S_y$ of the wave leaving the slit array oscillates while decreases; this is shown for $n=4.9$ in the inset of Fig.~\ref{fig:Sy-origgeom}. The effect is not so apparent for germanium partly because it is further from the critical value and partly because the material has nonzero conductivity ($\sigma=217$).

\begin{figure*}[tb]
\includegraphics[width=\textwidth]{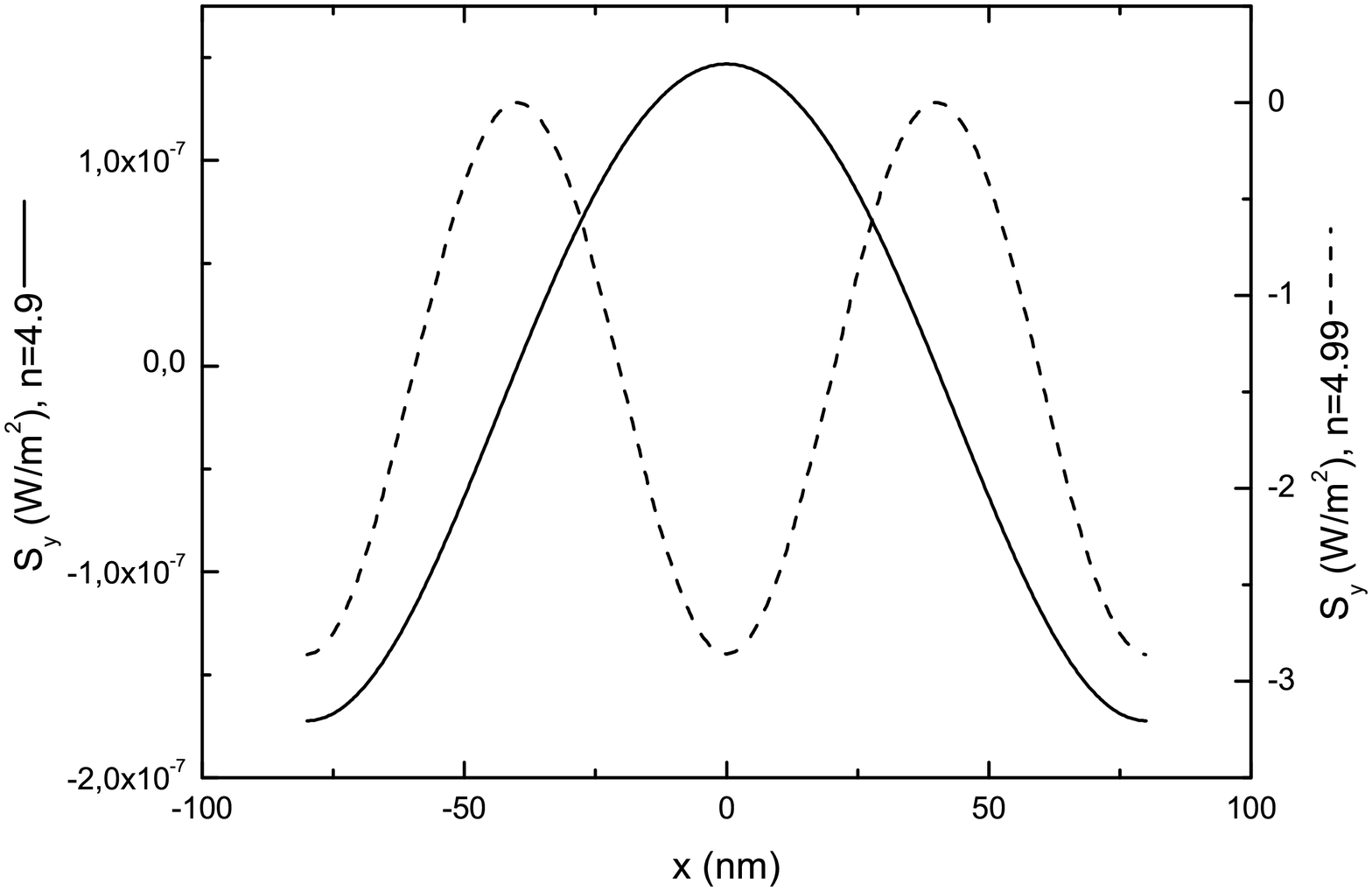}
\caption{\label{fig-origgeom-xmetsz}Energy flow $S_y$ along the $x$ direction at $y=32a$ for refractive indices $n=4.9$ and $n=4.99$. Parameters: wavelength $\lambda_0=800$~nm, slit width $2a=\lambda_0/10$, screen thickness $b=50$~nm. Transverse characteristic dimension of the geometry is $\Lambda=\left(\frac{4a}{2\pi}\right)$. Note that, although this is only an example of the beam shape at $y=32a$, our calculations showed, however, that the energy flow $S_y$ of the beam preserves its shape in the region $|y|>2a$ in both cases, i.e.\ this phenomenon cannot be attributed to self-imaging.}
\end{figure*}

To verify that the beam preserves its diffraction-free quality, in Fig.~\ref{fig-origgeom-xmetsz}
we plotted the energy flow $S_y$ along the $x$ direction at $y=32a$ for $n=4.9$ (evanescent region) and $n=4.99$ (nonevanescent region). We only present an example of the beam shape at $y=32a$, our calculations showed, however, that the energy flow $S_y$ of the beam preserves its shape in the region $|y|>2a$, i.e.\ this phenomenon cannot be attributed to self-imaging. Although separate $y$ axes must be assigned to the two cases because of the magnitude difference, the diffraction-free behavior can be observed in both cases; however, in the propagating case the originally single peak becomes double which shows a better localization.
This phenomenon, however, cannot be observed in the $E_x$ or $H_z$ components from which $S_y$ is calculated. The only difference that can be observed -- apart from the intensity increase -- is a change in the sign of the $E_x$ field. In the case of low refractive indices the sign of $E_x$ and $H_z$ are opposite while in the high-refractive-index case their sign is the same which results in a double peak.

\begin{figure}[tb]
\includegraphics[width=\columnwidth]{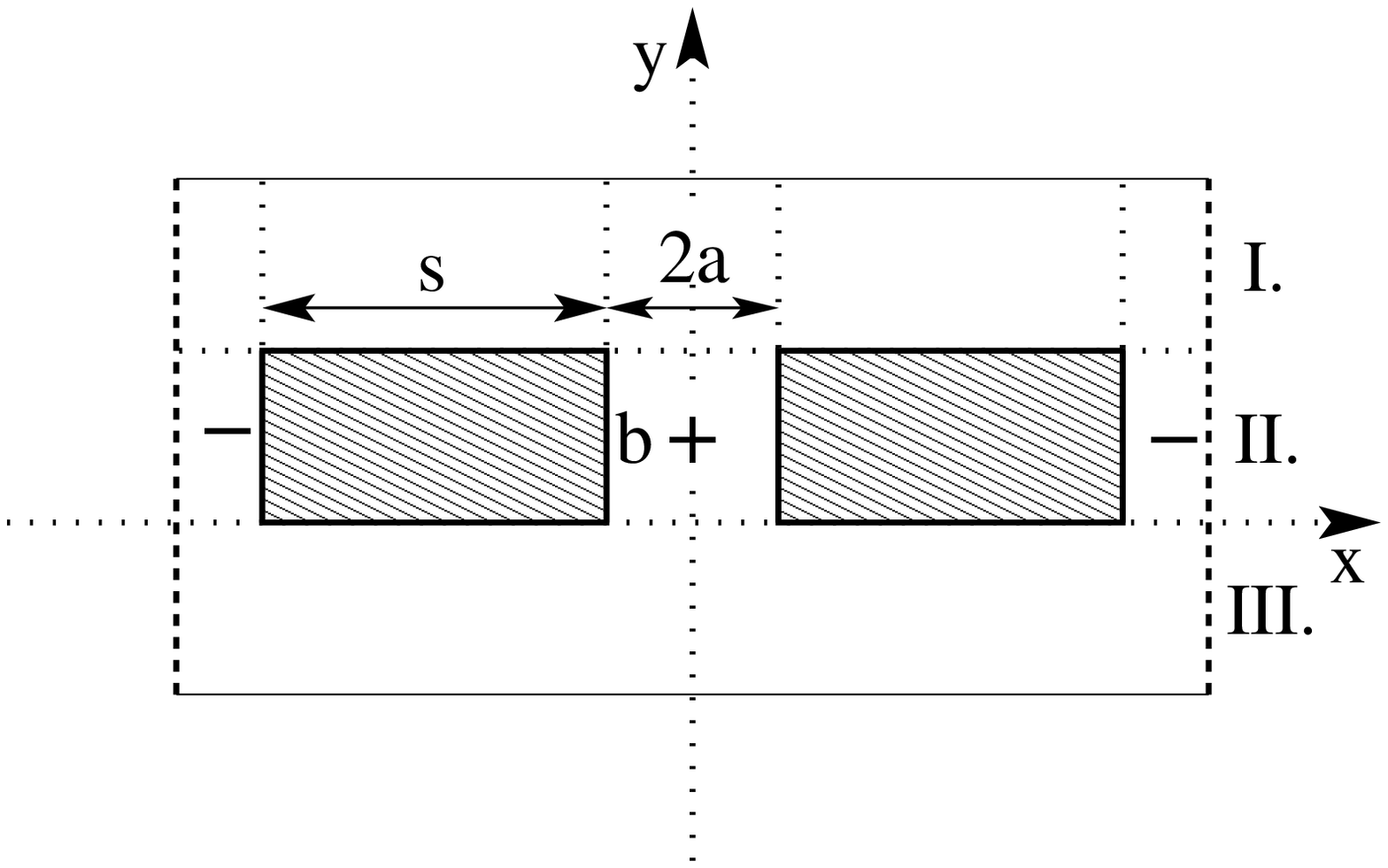}
\caption{\label{fig:newgeom}Modified geometry. The parameters are the same as in Fig.~\ref{fig:origgeom}, but the slits are separated by a feasibly thick wall (in our case the separation was $s=4a$).}
\end{figure}

\begin{figure*}[tb]
\includegraphics[width=\textwidth]{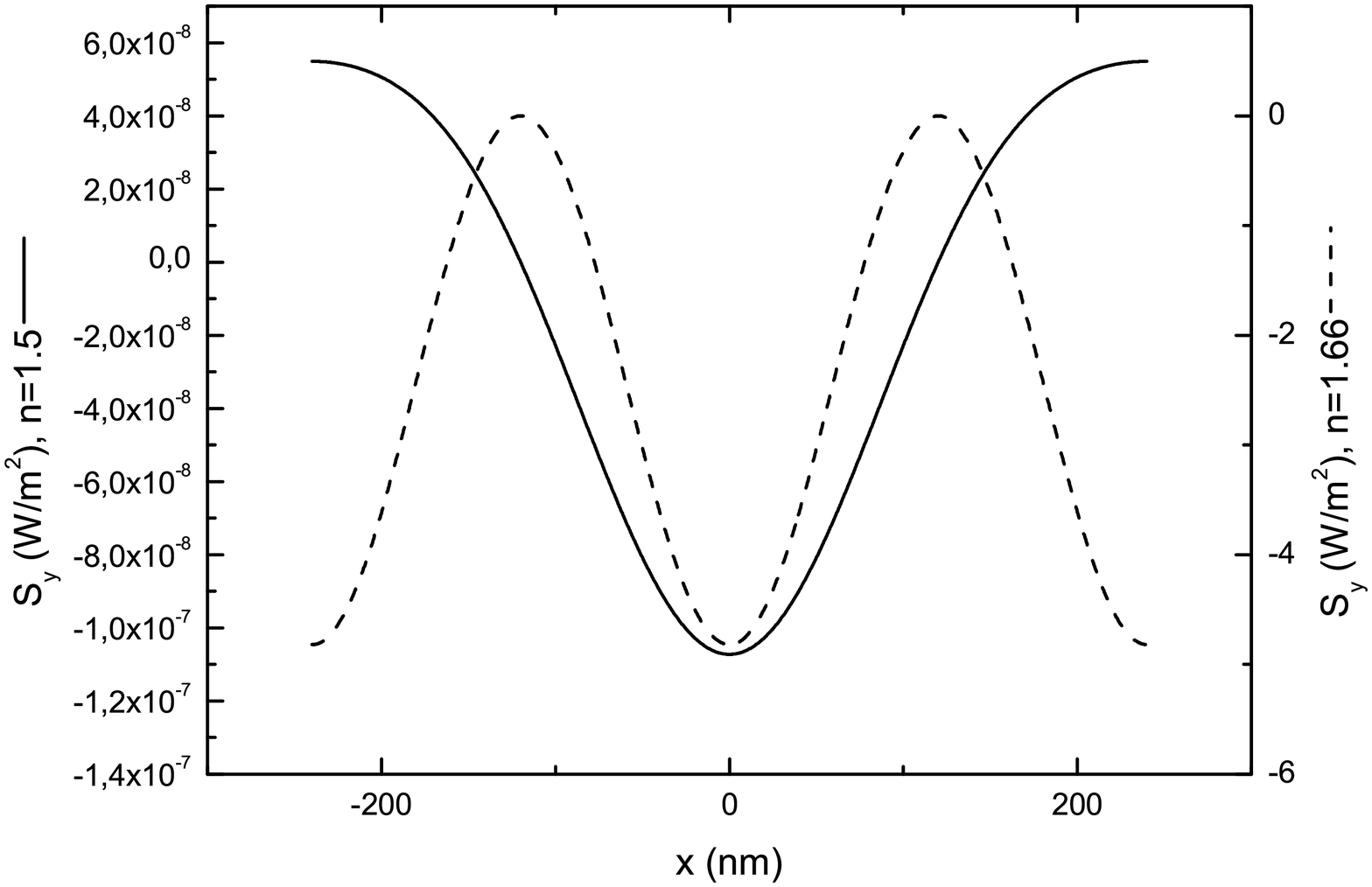}
\caption{\label{fig:newgeom-xmetsz}Energy flow $S_y$ along the $x$ direction at $y=32a$ for $n=1.5$ (evanescent region) and $n=1.66$ (nonevanescent region). Parameters: wavelength $\lambda_0=800$~nm, slit width $2a=\lambda_0/10$, screen thickness $b=50$~nm, separation $s=4a$. Characteristic dimension: $\Lambda=\frac{4a+2s}{2\pi}$, critical refractive index $n\approx 5/3$.}
\end{figure*}

In Figure~\ref{fig:newgeom}
we present a more realistic geometry. In this case the separation $s$ between the slits is nonzero. Actually, the separation can be arbitrary, and still the diffraction-free quality remains; as an example, in our simulations we used $s=4a$. The increase of the separation affects the characteristic dimension: it becomes $\Lambda=\frac{4a+2s}{2\pi}$. The width of the diffraction-free beams also changes: $w=2a+s$. In fact, the former value is the period of the slit system; the latter value is presented in Figure~\ref{fig:newgeom-xmetsz}.
In the figure we show the energy flow $S_y$ along the $x$ direction at $y=32a$ for $n=1.5$ (evanescent region) and $n=1.66$ (nonevanescent region). As it is expected from Eq.~\eqref{eq:ufg}, the numerical model showed that in order to produce propagating beams, increasing the characteristic dimension two times allows us to decrease the refractive index two times. It can be seen that, similarly to Fig.~\ref{fig-origgeom-xmetsz}, separate $y$ axes must be assigned to the two cases because of the magnitude difference, the diffraction-free quality of the beam still can be observed, and better localization can be observed for the propagating case.
 However, the width of the beam is much greater ($w=2a+s$) than in Fig.~\ref{fig-origgeom-xmetsz}, and the critical refractive index is reduced to $n\approx5/3$ which corresponds to a characteristic dimension of $\Lambda=\frac{4a+2s}{2\pi}$.

\begin{figure*}[tb]
\includegraphics[width=\textwidth]{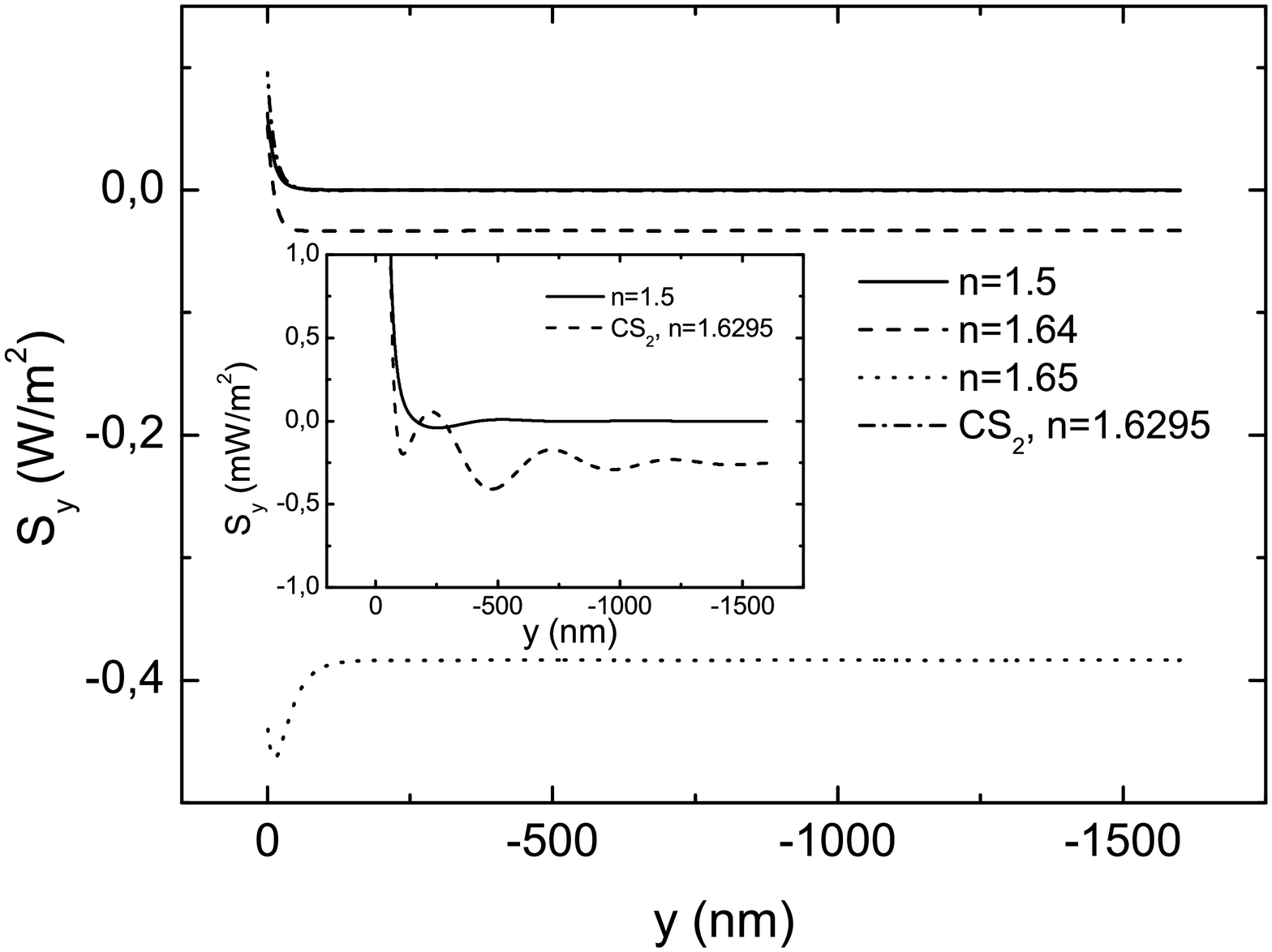}
\caption{\label{fig:newgeom-Sy}Energy flow $S_y$ along the $y$ axis at $x=0$ for refractive indices $n=1.5,1.64,1.65$. Parameters: wavelength $\lambda_0=800$~nm, slit width $2a=\lambda_0/10$, screen thickness $b=50$~nm, separation $s=4a$. Characteristic dimension: $\Lambda=\frac{4a+2s}{2\pi}$, critical refractive index $n\approx 5/3$. For comparison we plotted the same quantity for a real material with moderately high refractive index (carbon disulphide, $n\approx 1.6295$).}
\end{figure*}

For comparison we also show the energy flow $S_y$ along the $y$ axis at $x=0$ for various refractive indices (Fig.~\ref{fig:newgeom-Sy}).
In this case it is not only possible to find a real material with appropriate refractive index but these materials are liquid therefore they can be used in real experiments. As an example, in the simulation we used carbon disulphide ($n=1.6295$, $\sigma\approx0$ $@$ $\lambda=800$~nm). It is clear from the inset of Fig.~\ref{fig:newgeom-Sy} that while $S_y$ converges to zero for $n=1.5$, the value to which $S_y$ converges for CS$_2$ differs from zero, i.e.\ it is possible to realize diffraction-free nonevanescent beams with this geometry.


As a final remark we note that our results are in accordance with the Abbe-limit \eqref{eq:Abbe}. If $\sin\theta=1$, the expression changes to $d=\lambda/(2n)$. Substituting the values used for the simulations, we find that the Abbe-limit for $n=5/3$ is $d=240$~nm, and for $n=5$ it is $d=80$~nm. These are exactly the transverse characteristic sizes we used for the two cases. Consequently, there is a clear link between the evanescence limit and the Abbe-limit.

Our results showed that varying the refractive index of the material in which the diffraction-free beams are generated leads to nonevanescent beams. These nondiffracting beams show some unique properties useful for both physical and technical applications
. Nondiffracting beams may be found applications in imaging as it wa verified that extremely long focal depths can be produced in imaging realized with such beams\cite{Bouchal66}. Particle acceleration is another field of research in which diffraction-free beams may be used; these beams can serve as alternative accelerating fields owing to the strong longitudinal component of the electric field \cite{Bouchal72}. It is also possible to manipulate ensembles of particles in multiple planes \cite{Bouchal73, Bouchal74} which can be realized due to the self-reconstruction ability of the nondiffracting beams. As a final example, we draw attention to the possibility of realizing light motors driven by the diffraction-free beam. However, extracting this nondiffracting nonevanescent beam from the higher refractive index material to air or any lower refractive index one is a challenging problem.
\section{Conclusion\label{sec:conclusion}}
In our paper we show that the production of propagating (nonevanescent) diffraction-free nano-beams using the Fresnel-waveguide concept is feasible. The analytical expression derived for nonevanescent beams in Section~\ref{sec:theory}, similarly to the well-known Abbe-limit, connects the wavelength in vacuum, the refractive index and the characteristic size of the slit system. We presented the results of finite element simulations for two slightly different geometries and various refractive indices as well as real materials. Our numerical results support the findings presented in Section~\ref{sec:theory}.
\begin{acknowledgments}
\add{The authors gratefully acknowledge Prof.\ Shlomo Ruschin for his helpful comments and suggestions.}
This work was supported in part by the Hungarian ELI project (\mbox{hELIos}, ELI\_09-1-2010-0013, ELIPSZTE), and the Hungarian "Social Renewal Operational Programme" (T\'AMOP 4.2.1./B-10/2/KONV-2010-0002).
\end{acknowledgments}

\end{document}